\documentclass[aps,prb,twocolumn]{revtex4-1}
\usepackage[english]{babel}
\usepackage[utf8]{inputenc}
\usepackage{amsmath}
\usepackage{amssymb}
\usepackage[caption=false]{subfig}
\usepackage{amssymb}
\usepackage{epsfig}
\usepackage{graphicx}
\usepackage{amsmath}
\usepackage{array,color}
\usepackage{natbib}
\usepackage{soul}
\usepackage[usenames,dvipsnames]{xcolor}
\usepackage{color, colortbl}
\definecolor{forestgreen}{rgb}{0.11,0.54,0.15}
\definecolor{purple}{rgb}{0.62,0.10,0.96}
\definecolor{dockerblue}{rgb}{0.11,0.56,0.98}
\definecolor{freeblue}{rgb}{0.25,0.41,0.88}
\usepackage[pdftex,plainpages=false,colorlinks=true,linkcolor=Red, citecolor=blue, urlcolor=blue]{hyperref}
\begin{document}
\title{Coupling strongly correlated electron systems to a tunable electronic reservoir}
\author{S. Bag$^{1,+}$, L. Fratino$^{1}$, A. Camjayi$^2$, M. Civelli$^1$, M. Rozenberg$^{1}$}
%
\affiliation{$^1$Universit\'e  Paris-Saclay, CNRS Laboratoire de Physique des Solides, 91405, Orsay, France.}

\affiliation{$^2$ Universidad de Buenos Aires, Ciclo B\'asico Com\'un and CONICET - Universidad de Buenos Aires, Instituto de F\'isica de Buenos Aires (IFIBA), Buenos Aires, Argentina.}
\affiliation{$^+$Department of Physics, Arizona State University, Tempe, AZ 85287, USA}

\date{\today}
\begin{abstract}
 We study the effect of coupling an electronic reservoir to a Hubbard model and to a Dimer Hubbard Model.
 This is motivated by recent experiments on the effect of illumination on the insulator-metal transition in a vanadium oxides and photo-conductive cadmium sulfide heterostructure.
We model the system as an electronic reservoir hybridized to the correlated system. We assume that the light intensity controls the hybridization coupling strength.
We find that the light intensity acts similarly as the temperature in the weak interaction regime. 
This is consistent with the role played by electronic 
reservoirs in out-of-equilibrium systems. 
In contrast, qualitative differences appear at strong coupling.
We show that modeling the V$_2$O$_3$ compound with a
Hubbard model, our results describe qualitatively
well the observed illumination-driven suppression of the insulator-metal transition. 
In contrast, in the DHM results fail to capture the
mild suppression observed in the case of VO$_2$.
This indicates 
that the lattice may play 
an important role in this case. 
\end{abstract}
\maketitle

\section{Introduction}\label{introduction} 

Mott insulator materials are interesting for neuromorphic devices, such as artificial spiking neurons\cite{delValleJAP, Stoliar2017, WilliamsNeuristor}. The several 
orders-of-magnitude changes of the resistivity across the 
metal-to-insulator-transition (MIT) of Mott insulators 
can be exploited in artificial neuromorphic devices to recreate the all-or-nothing excitation of an action potential \cite{Stoliar2017,WilliamsNeuristor,luo2022spin,qiu2023stochasticity,Shaobo}. 
Furthermore, there are several ways to control the Mott MIT, 
namely by changing doping, pressure, electric field, etc. 
\cite{kumar2013local,chen2017sequential,yang2016suppression,schlom2014elastic,aetukuri2013control,matsuda2020magnetic,wu2011electric,jeong2013suppression,wan2018limiting,dong2018lithography,navarro2021hybrid,ke2018vanadium,singer2018nonequilibrium}. 
A particularly interesting possibility, in the context of neuromorphic applications, is to control the
transition by illumination with light in hybrid thin-film structures where one changes the coupling between the Mott material and its environment \cite{navarro2021hybrid,navarro2023light,adda2022optoelectronic}

 In recent years, the study of phenomena that emerge from the control of the 
 coupling between a system to a reservoir is attracting significant attention. 
 In experiments, changing the environment coupling can lead, for instance, to line-width broadening\cite{line_brodening}, decoherence, and a finite lifetime of states\cite{decoherence_and_lifetime}. 
 Moreover, it can allow tailoring desired states, like  entangled\cite{entangled}, antiferromagnetically (AFM) ordered\cite{AFM_dissipation}, Bose condensed \cite{Quantum_states_drive}, etc. While this has been so far considered 
 mostly in the context of cold atoms, it is also potentially relevant for electronic
 properties of solid state systems, including strongly correlated ones. For instance,
 an important case is that of systems with metal-insulator transitions that are
 driven out of equilibrium. In those cases, it has been recognized that 
 coupling to an electronic reservoir is an essential physical ingredient
 to achieve a steady state in a static electric field-driven Hubbard 
 Model (HM) \cite{Amaricci,cami1,javier1,javier2,dienner,rodolfo}. 
 Hence, it is an important question to consider the systematic effect that 
 the coupling to an electronic reservoir may have on a given strongly correlated 
 system.



In addition to the previous motivation, the effect of a (semi)metallic 
reservoir coupled to a strongly correlated system is also relevant for
hetero-structures. Indeed, the ability to grow multi-layers
including high quality strongly correlated oxides, 
prompted the interest in understanding the fate of the Mott metal-insulator 
transition in metal and semi-metal / Mott-insulator hybrid systems
\cite{thiel}. 

For example, in the study of hetero-structures of a metal/AFM-Mott  model, a
suppression of AFM structure factor is observed in the AFM-Mott layer due to 
its proximity to the 
metal\cite{Hm-M3, Kondo-Heisenberg2}. However, it was also reported that
the effect of an additional conduction band in the periodic Anderson model can
lead to stabilization of the AF order\cite{Scalettar, Vidhyadhiraja, KI-M} in the system. 
Furthermore, in the hetero-structure of a paramagnetic (PM) metal and a PM-Mott 
insulator, the metal state may penetrate into the Mott insulator side \cite{HM-M1, HM-M2,zenia}.

A particularly interesting type of Mott insulator materials are the vanadium 
oxides VO$_2$ and V$_2$O$_3$, which have temperature-driven MITs \cite{IFT}.

In a recent experiment, the effects on the MIT was studied in the heterostructure of 
vanadium oxide and the photoconductive semiconductor Cadmium Sulfide (CdS), for
various levels of illumination \cite{navarro2021hybrid}. 
Interestingly, both VO$_2$ and V$_2$O$_3$ showed a suppression of MIT with the power 
of light illuminated on CdS. However, the modulation of the MITs in VO$_2$ and V$_2$O$_3$ 
were dramatically different. 
The illumination quickly suppressed the $T_{\rm MIT}$ of V$_2$O$_3$, driving it 
down toward zero temperature. 
While it only had a minor suppression effect on  $T_{\rm MIT}$ of VO$_2$, shifting
it down by just a few degrees.

The present study is motivated by those experiments. We aim to understand the suppression of $T_{\rm MIT}$ in Vanadium Oxides/CdS heterostructures in the presence of light. 
To that goal, we need to model both the effect of the CdS and also the MITs in V$_2$O$_3$ and VO$_2$.
In the case of V$_2$O$_3$, the MIT occurs between an AFM insulator to a PM metallic state. 
In this initial study, for the sake of simplicity, we shall follow the recent study \cite{V2O3_HM} to model the transition. In that work, it was shown that the  
single-band HM treated within DMFT was able to capture non-trivial features 
of the T driven MIT with AFM symmetry breaking at half-filling, such as
the anomalous enhancement of the negative magneto-resistance.
On the other hand, the MIT in the case of VO$_2$ is qualitatively different, since 
it is non-magnetic.
Moreover, the Mott insulator phase of VO$_2$ is monoclinic, where pairs of V 
atoms are dimerized, presumably forming "dynamical singlets" \cite{biermann_PRL2005}. 
Fortunately, in this case we can also count on a simple model that captures basic 
non-trivial features of the MIT, which we can adopt in this initial study.
This is the Dimer Hubbard Model (DHM), which is
an extension of the Hubbard model where each unit cell contains a dimer.
It was recently shown that the DMFT solution of the model exhibits a 
T-driven first order non-magnetic MIT at half-filling, 
where the ground-state is a dimerized Mott insulator with dynamical singlets 
\cite{Oscar2018,ExpVO2,ocampo2017study,Oscar2017}. 
Hence, in this initial study we shall adopt these two simple models 
as zero-order approximation to describe the MITs of V$_2$O$_3$ and VO$_2$
in the presence of a tunable hybridization.

%




%

One of our main results is that the reservoir 
may qualitatively act as a temperature and
can may suppress the T$_{MIT}$ in
agreement with the experimental observations
on V$_2$O$_3$. 
However, the comparison with
observations in VO$_2$ there seems to
remain some significant quantitative 
differences. We shall argue later on that
this may point to a relevant role played
by the lattice degree of freedom that are
not included in our model \cite{MFAA}.


This paper is organized as follows: In Section~\ref{Method} we describe the model 
and the numerical approaches. 
Then, in Section~\ref{Result} we shall describe the results. 
Firstly, we discuss the systematic effect of $T$ 
and $\Gamma$ on the AFM phase and the MIT of the single band HM. 
Secondly, we shall describe 
the effects of those parameters for the interaction-driven PM insulator-metal transition 
in both the HM and the DHM. While the former is not directly relevant to the physics of 
the vanadates, we find it useful to include that study as a reference case, 
since it is the most widely studied MIT within DMFT. 
In the last Section~\ref{ConclusionAndDiscussion}, 
we discuss the comparison of our findings with experimental results, including 
the possible origin of some standing discrepancies.

\section{Models and Method}\label{Method}

Our task is to model the hybridization effect of CdS, which is a good photoconductive semiconductor\cite{CDS1,CDS2}. 
In dark it is highly insulating, i.e. more than the Mott insulator states in the vanadium oxides. However, when illuminated by light it creates a substantial amount of electron-hole pairs and, in consequence, shows a drop in the resistivity. 
Nevertheless, CdS never has truly metallic character but remains a semiconductor, 
i.e. does not have metallic conduction. 
Therefore, we propose to model the CdS as a set of electronic reservoirs that 
that are locally coupled with the V sites. In other words, we consider 
that the excited electron-hole pairs are described as incoherent metallic
states that locally hybridize with the vanadium oxide's. 
Since the vanadium oxides are thin films of the order of 10 nm, 
we shall also assume that the electric conduction is dominated by the physics of the
interface, which we model as a layer of HM or DHM hybridized with electronic reservoirs 
at each lattice site.
Since increasing the light intensity on the CdS one can tune the number of (photo)carriers,
we model this effect through the strength of the hybridization 
parameter $\Gamma = \gamma^2 \rho(E_F)$. 
$\gamma$ is a geometrical parameter independent of the light intensity as it describes the hybridization hopping amplitude between the sites that represent the 
CdS and the vanadium oxide layers, respectively. On the other hand, 
$\rho(E_F)$ represents the density of states of the CdS, which increases with light intensity. However, the detailed dependence of $\rho(E_F)$ and the
intensity of illumination, that is experimentally controlled by the 
power on the LED light, is not trivial. Therefore, here
for the sake of definiteness and simplicity, we assume 
that the control parameter $\Gamma$ (through the effect of $\rho$) 
is proportional to the power of the LED.

\subsection{Hubbard Model (HM)}
\begin{figure}[h!]
\begin{center}
\includegraphics[width=1.1\linewidth]{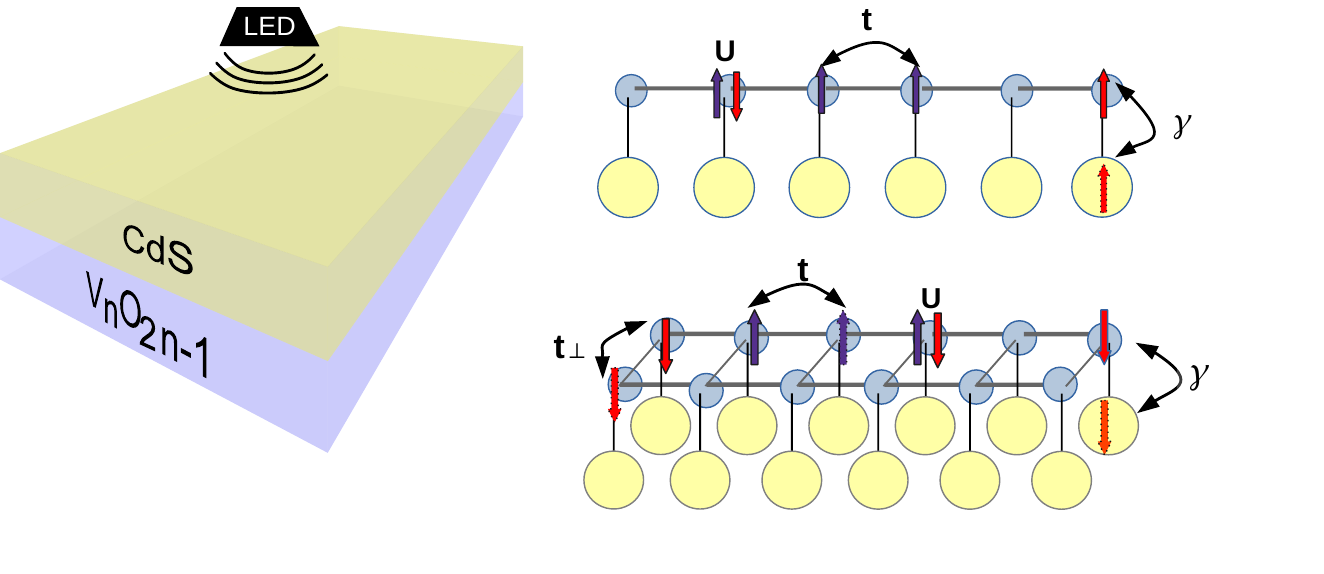}
\end{center}
\caption{schematic experimental setup for CdS/V$_n$O$_{2n-1}$ heterostructures with light on CdS where $n=2$ corresponds to CdS/V$_2$O$_3$ and  $n=\infty$ corresponds to CdS/VO$_2$. Schematic representation of HM (top-right) and DHM (bottom-right) with electronic reservoir connected to each site of the lattice. $t$ is the hopping amplitude of an electron between nearest-neighbor sites. $\gamma$ is the hopping amplitude of an electron between the site and the electronic reservoir. $t_\perp$ is the hopping amplitude of an electron between dimer sites.}\label{sche_EXP_HM_DHM}
\end{figure}

The Hamiltonian of the single-band HM coupled with an 
incoherent electronic reservoir reads as
\begin{align}\nonumber
	\hat{\mathcal{H}} =& -t\sum_{\langle i,j \rangle, \sigma}( \hat{c}^\dagger_{i \sigma}\hat{c}_{j \sigma}  + h.c)\\ 
	&+U\sum_{i}\hat{n}_{i \uparrow}\hat{n}_{i \downarrow}-\mu \sum_{i,\sigma}\hat{c}^\dagger_{i \sigma} \hat{c}_{i, \sigma} + \sum_{i,\sigma} \hat{\mathcal{H}}^{i\sigma}_{c},
\label{Method:Hubbard_model}
\end{align}
where $\hat{c}^\dagger_{i \sigma}$ and $\hat{c}_{i \sigma}$ are respectively the fermionic creation and annihilation operators at the $i$-th site of the lattice and  $\hat{n}_{i\sigma} = \hat{c}^{\dagger}_{i\sigma}\hat{c}_{i\sigma}$ is the number operator. Here $t$ is the hopping amplitude of an electron between nearest-neighbor sites, $U$ is the on-site repulsion energy if two electrons (of opposite spins) occupy the site, and $\mu$ is the chemical potential of the system, controlling filling. We fixed $\mu$ at $U/2$ for the half-filled case. Last term of the Hamiltonian, $\sum_{\sigma}\hat{\mathcal{H}}^{i\sigma}_{c}$, represents the coupling of an electronic reservoir with $i$-th site of the lattice. Details about this coupling term 
are given in the later part of this section. 
The top-left panel of Fig.~\ref{sche_EXP_HM_DHM} displays a schematic 
representation of HM with the electronic reservoir connected to each site of the lattice.
\\

We solve this model using DMFT with a semicircular density of states:
 $D(\epsilon) =\frac{2}{\pi D^2}\sqrt{(D^2- \epsilon^2)}$ with half band width $D=1$. DMFT provides the exact solution of the model\cite{DMFTRMP,V2O3_HM} on the Bethe lattice. 
 We solve the impurity problem of DMFT using Hybridization Expansion-Continuous-Time Quantum Monte Carlo (HYB-CTQMC) \cite{rmpqmc,WernerCTQMC,ctqmchaule}. 
 We solve the model both with and without AFM symmetry. We present the result of the HYB-CTQMC in the main text. 
 
\subsection{Dimer Hubbard Model (DHM)}
An interesting extension of single-band HM is the DHM \cite{goetz}. 
This schematic model was shown to capture some key qualitative features of the temperature-driven MIT transition of VO$_2$ \cite{Oscar2017,Oscar2018,stinson2018imaging}. 
The DHM Hamiltonian, coupled with an incoherent electronic reservoir reads,
\begin{align}\nonumber
\hat{\mathcal{H}}&= -t\sum_{\langle i, j\rangle , \sigma,\alpha} \left(\hat{c}^\dagger_{i\alpha\sigma} \, \hat{c}_{j\alpha\sigma}+ \mathrm{h.c}\right)+\\\nonumber
&+t_\perp\sum_{ i, \sigma} \left(\hat{c}^\dagger_{i1\sigma} \, \hat{c}_{i2\sigma}+ \mathrm{h.c}\right) +\\
&+ U \sum_{i,\alpha} \hat{n}_{i\alpha\uparrow} \, \hat{n}_{i\alpha\downarrow} - \mu \sum_{i,\alpha,\sigma} \hat{n}_{i \alpha \sigma}+ \sum_{i,\alpha,\sigma} \hat{\mathcal{H}}^{i\alpha\sigma}_{c},
\label{DHM}
\end{align}
where the index $i$ and $j$ denote the lattice cells, $\alpha=1,2$ denotes the dimer sites within a given cell and  $\sigma$ labels the spin.  
The hopping $t$ and $t_\perp$ correspond to the amplitudes between nearest-neighbor lattice cells and between dimer sites, respectively. 
The last term of the Hamiltonian, $\sum_{\sigma}\hat{\mathcal{H}}^{i\sigma\alpha}_{c}$, represent the electronic reservoir coupled to $i$-th lattice cell's dimer site $\alpha$. Note that we use an independent electronic reservoir for each site of the dimer as 
can be seen in the schematic representation of the model in the bottom-right panel of Fig.~\ref{sche_EXP_HM_DHM}.
Details about the electronic reservoir are given later in this section.  
The local Green function of the lattice becomes diagonal in the Bond ($B$) and Anti-Bond ($AB$) basis, 
where the creation operator in the $B$ and $AB$ basis is related to the site basis by the following equation:
\begin{align}
 \hat{c}^\dagger_{iAB/B\sigma}= \frac{\hat{c}^\dagger_{i1 \sigma} \pm \hat{c}^\dagger_{i2\sigma}}{\sqrt{2}}
 \end{align}
Therefore, we find it practical to solve the model on this basis. 



\subsection{Electronic Reservoirs}
The Hamiltonian of the electronic reservoir coupled to each lattice site of the models (HM or DHM) reads,
\begin{equation}
\sum_{\sigma} \hat{\mathcal{H}}^{i\sigma}_{c} = \sum_{k,\sigma}\epsilon_k \hat{a}^\dagger_{ik\sigma} \hat{a}_{ik\sigma} + \gamma \sum_{l,\sigma}\left( \hat{a}^\dagger_{il\sigma} \hat{c}_{i\sigma} + h.c\right)
\label{Metallic}
\end{equation}
 where $\hat{c}_{i\sigma}$ denotes the fermionic operator at site $i$ for the HM (or similarly, adding a label $\alpha$, 
 at the $i$-th cell's $\alpha$ site for DHM) 
 and $\hat{a}_{il\sigma}$ is the fermionic operator for reservoir's electron. $\epsilon_l$ are the reservoir's electron energy levels. 
 For simplicity, we assume that the reservoir's electrons hybridize with the lattice fermions with constant
 amplitude $\gamma$. We can integrate out the non-interacting reservoir's electrons, and their effect appears 
 as an additional effective hybridization for the $c$ fermions at each impurity site, which remains fixed in the DMFT calculation. 
This effective hybridization due to the reservoir's electrons reads as
\begin{equation}
 \Delta_{\Gamma}(\omega_n)=  \sum_{l} \frac{\gamma^2}{i\omega_n-\epsilon_l}=\int d\epsilon \rho_l(\epsilon)\frac{\gamma^2}{i\omega_n-\epsilon}
 \end{equation}
For simplicity, we adopt a semi-circular Density of States for the reservoir electrons,
$\rho_l(\epsilon)=\frac{2}{\pi D'^2}\sqrt{(D'^2- \epsilon^2)}$, so the above integration takes 
the closed form
\begin{equation}
\Delta_{\Gamma}(\omega_n) =    - \frac{2D'i\Gamma}{\omega_n + sig(\omega_n)\sqrt{\omega_n^2+D'^2}}
\label{Sigmac}
\end{equation}
where $\Gamma = \gamma^2/D'$ controls the strength of the effect of hybridization with a reservoir.
For convenience, we choose half the bandwidth of the electronic reservoir $D'=1.0$. 

\subsection{Details on the DMFT method} 
We describe here how the hybridization of the associated impurity problem is modified by the presence of the reservoirs.

In the case of the HM we shall consider both solutions with and without AFM symmetry. The effect of the reservoir within DMFT is most clearly observed through the self-consistency condition of the associated
quantum impurity problem \cite{DMFTRMP}. 
The total hybridization function reads, 
\begin{equation}
    \Delta_{\sigma}(\omega)=t^2 G_{loc,\bar{\sigma}}(\omega)+\Delta_{\Gamma}(\omega)
        \label{self-cons}
\end{equation} 
where we observe that the reservoir is simply
added to the quantum single site environment.
For further details on the DMFT method and
self-consistent equations for the AFM-HM see 
Refs.\onlinecite{DMFTRMP,V2O3_HM}.

In the case of the DHM, we consider the $AB/B$ basis, hence the total hybridization entering the DMFT self-consistency equation 
becomes 
\begin{equation}
    \Delta_{AB/B\sigma}(\omega)=t^2 G_{loc, AB/B\sigma}(\omega)+\Delta_{\Gamma}(\omega)
    \label{self-cons2}
\end{equation} 
For further details
on the DMFT self-consistent equations for the DHM see Refs.\onlinecite{Oscar2017,Oscar2018,LorenzoSoumen}.

As model parameters, we adopt in the case of the HM $t=0.5$ (since we already set $D=1$), and 
we fix onsite local repulsion $U=1.7$, which was found adequate to describe the AFM-PM MIT in V$_2$O$_3$\cite{V2O3_HM}. 
For the DHM we also adopt $t=0.5$, and then fix $t_\perp=0.3$ and the onsite local repulsion $U=2.2$, as where already
shown to be adequate to qualitatively describe the PM MIT in VO$_2$\cite{Oscar2017,Oscar2018}.
In addition, we shall also consider for both models and for the sake of obtaining a full qualitative
picture of the behavior, a large value of $U$=4. This sets the systems well into their Mott insulator states,
so the MIT actually would become an ``insulator-insulator" transition. However, as we shall see, the
presence of the reservoirs will change this naif expectation.


The DMFT calculation with the HYB-CTQMC impurity solver produces data in Matsubara Frequency. However,
to obtain the behavior of the density of state (DOS) and the DC resistivity, one requires real-frequency data. 
Therefore, we use the maximum entropy method to get the analytical continued Green's function\cite{ctqmchaule,MEM}. 
From the Green's function in real frequency, one can then calculate the local DOS ($\mathcal{A}(\omega)$), which 
reads
\begin{align}
\mathcal{A}(\omega)= -\frac{1}{\pi} Im \,G(\omega^+)
\end{align}

For the HM we shall show the total DOS, i.e. the average of the two spin projections, 
$\mathcal{A}_{HM}(\omega)=\frac{1}{2}\sum_{\sigma}\mathcal{A}_{\sigma}(\omega)$. 
Whereas, for the DHM, the total DOS results from the average of spin and B/AB-symmetry projections, 
$\mathcal{A}_{DHM}(\omega)=\frac{1}{4}\sum_{\sigma}\left( \mathcal{A}_{B,\sigma}(\omega)+\mathcal{A}_{AB,\sigma}(\omega)\right)$. 
Expressions for the conductivity of the HM and the DHM are given in the supplementary material. 

\section{Results and Discussion}\label{Result}

\subsection{Hubbard Model with antiferromagnetic symmetry}
\begin{figure}[h!]
\begin{center}
\includegraphics[width=1.0\linewidth]{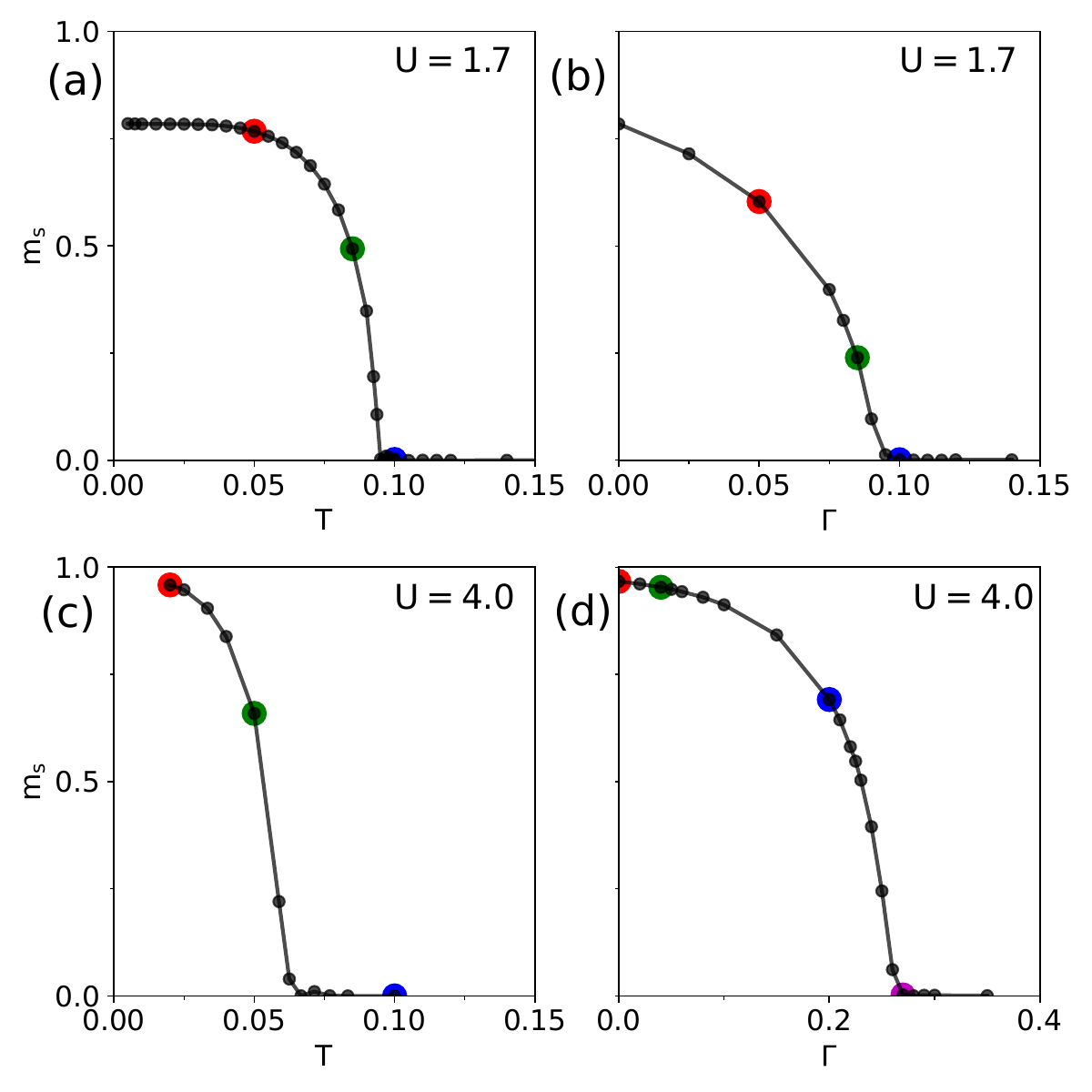}
\end{center}
\caption{(a) and (b) the variation of the staggered magnetization as a function of $T$ and $\Gamma$, 
respectively, for HM at $U=1.7$. (c) and (d) the same at $U=4.0$\label{MS_T_Gamma}.
We observe that both thermal fluctuations and $\Gamma$ can suppress the AFM order, seemingly
playing a qualitatively similar role ($T$=0.01 in panels (b) and (d)).}
\end{figure}

We begin with the MIT in the single-band HM model case, which occurs in the presence of antiferromagnetism.
In the absence of an electronic bath, the low-temperature state is an AFM insulator, and there is a $U$-dependent
$T_N$ where magnetization goes to zero, and the gap closes, and one has the $T$-driven MIT. The presence of an AFM state is signaled by
the staggered magnetization ($m_s = |n_\uparrow -n_\downarrow|$) order parameter.

\begin{figure}[h!]
\begin{center}
\includegraphics[width=1.0\linewidth]{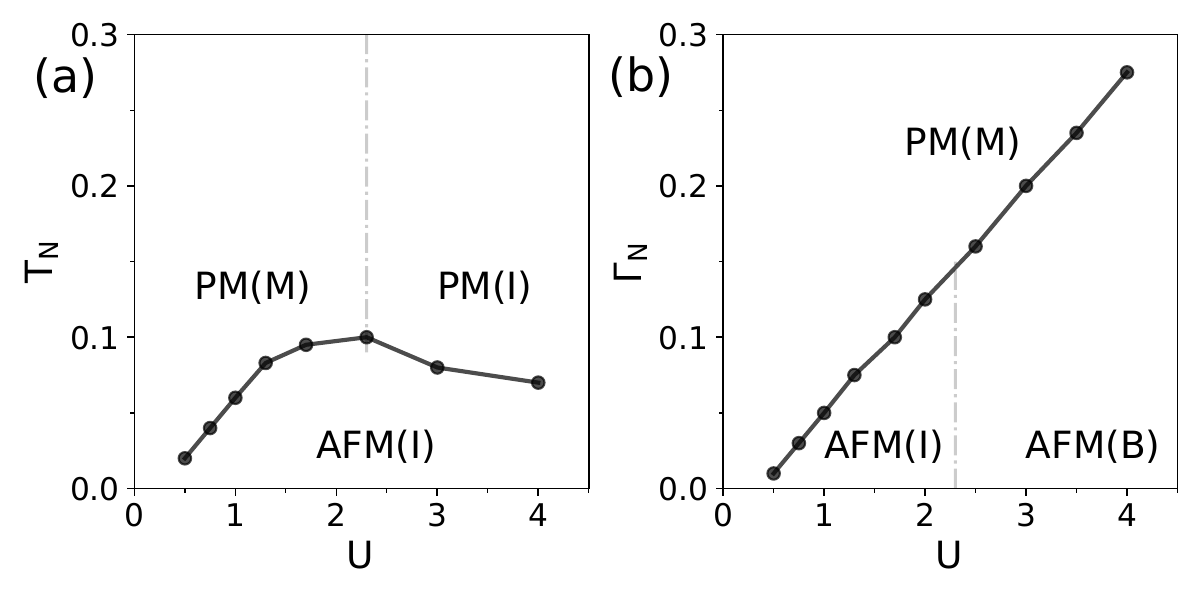}
\end{center}
\caption{(a) $T$-$U$ phase diagram for $\Gamma=0.0$ and (b) $\Gamma$-$U$ phase diagram for $T=0.01$ of the half-filled HM 
with AFM symmetry. 
M, I, and B label Metal, Insulator, and Bad-metal, respectively. 
Dashed-lines approximately denote the phase boundaries.}\label{HM_AFM_phase_dia}
\end{figure}

Fig.\ref{MS_T_Gamma} shows how $m_s$ continuously goes to zero as a function of $T$ for $U=1.7$ and for
 $U=4.0$, when $\Gamma$=0. 
 Interestingly,  similarly as the effect of $T$, we observe that $\Gamma$ can also suppress the AFM ordering in HM 
 at both low and high $U$, when $T$ is fixed at a small value. 
One may be tempted to conclude that $\Gamma$ behaves qualitatively as $T$. 
However, the systematic results shown in Fig.~\ref{HM_AFM_phase_dia} shows otherwise.
They display the variation of Neel's temperature ($T_N$) and, similarly the $\Gamma_{\rm N}$ that
signals the critical $\Gamma$ value that drives the magnetization to zero, as a function of $U$.
The data reveals that the dependence is qualitatively different.
In the absence of reservoirs, $T_N$ has the well-known non-monotonous dependence on $U$\cite{DMFTRMP,soumen,Lorenzo}. 
For small $U$, $T_N$ increases with increasing $U$, as the AFM gap increases with both $m_s$ and $U$.
While at large $U$, $T_N$ changes its behavior and decreases with increasing $U$, since the magnetic
exchange coupling decreases as $\sim t^2/U$.
For the case of $\Gamma$ at low $T$ we observe that for low values of interaction $U$, 
it seems to produce an effect similar to $T$, as was expected from the results of Fig.\ref{MS_T_Gamma}. 
However, in stark contrast, we observe that at larger values of $U$, the critical $\Gamma_{\rm N}$ 
increases monotonously with the interaction strength. This means that the reservoir is no
longer as effective in suppressing the magnetism.

The physics driving the suppression of $m_s$ with increasing in $\Gamma$ is qualitatively different
from the thermal disorder induced by $T$.
It can be attributed to the Kondo screening of magnetic moment at each site of HM due to the 
electronic reservoir. This can be most clearly seen if, for a moment, we set the hopping $t$
of the HM to zero, and we obtain a collection of single Anderson impurity models (AIM), at every site. 
The bath of the AIM is controlled by the strength of $\Gamma$.
Thus, the local moment of the lattice sites are induced by $U$, but screened by $\Gamma$
through the Kondo effect.  
Therefore, with increasing $U$, one needs a larger $\Gamma_N$ to screen $m_S$ and suppress the
magnetic order. 

\begin{figure}[h!]
\begin{center}
\includegraphics[width=1.0\linewidth]{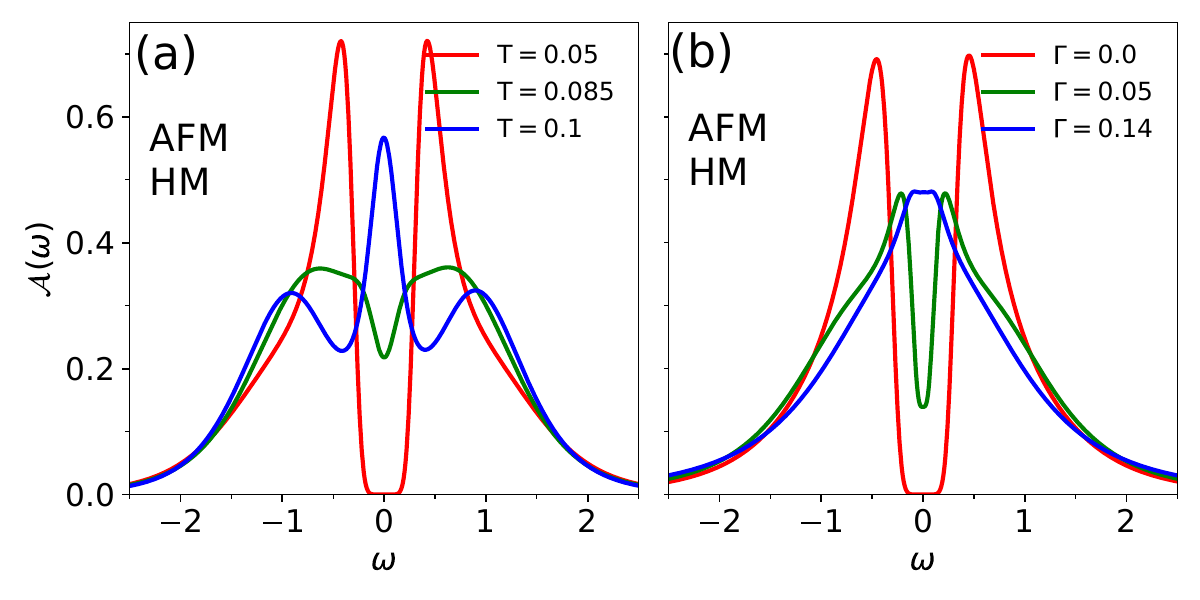}
\end{center}
\caption{Comparison of the evolution of the total (i.e. spin averaged) local DOS 
as a function of $T$ at fixed $\Gamma=0.0$ (a)
and of $\Gamma$ at fixed $T=0.01$ (b), at interaction $U$=1.7, for the AFM-HM across the AFM to PM. 
The line colors correspond to the spots along the $m_s$ curve of Fig.\ref{MS_T_Gamma}.}\label{DOS_AFM_U1.7}
\end{figure}

To gain a deeper understanding of the observed contrast between similar behaviors of $\Gamma_N$ and $T_N$ 
at small $U$ but different at large $U$, we study the evolution of the local DOS. We have
selected specific points across the $T-$ and $\Gamma-$driven transitions indicated by the color 
spots in Fig.\ref{MS_T_Gamma}. The panel (a) and (b) of 
Fig.\ref{DOS_AFM_U1.7} displays the variation of the local DOS with $T$ and $\Gamma$, respectively, for $U=1.7$. 
We recall that the DOS looks symmetric because we are showing the total DOS, 
which is the sum of the up and down spin components, since we are mostly interested in its
qualitative evolution.
At small $T$, the HM has AFM order and, consequently, the DOS has a gap at the Fermi energy. 
As one increases $T$, $m_s$ decreases, and the gap gradually decreases and fills up. 
For large enough $T$, $T \geq 0.09$, $m_s$ vanishes and the DOS develops a prominent 
quasi-particle peak at Fermi energy. This is the familiar Kondo-peak that develops in the
associated impurity problem \cite{DMFTRMP}. 
A qualitatively similar behavior is observed in the variation of DOS as one 
increases $\Gamma$ while keeping $T$ fixed. However, some quantitative differences are
apparent. For instance, the Hubbard side-bands are not so well resolved as the 
system becomes metallic. This can be understood from the fact that the metallic
quasi-particle peak is the manifestation of the Kondo peak in the associated impurity 
problem. In contrast to the previous $\Gamma$=0 case, now, the finite value
of $\Gamma$ implies
that there is a non-renormalizable component of the impurity bath which does not change
under the self-consistency condition \ref{self-cons}. This ``wide'' hybridization 
increases as $\Gamma$ increases, driving the width of the Kondo peak larger and larger.
Hence, the quasi-particle peak is wider, taking significant spectral weight from the Hubbard 
band components.

\begin{figure}[h!]
\begin{center}
\includegraphics[width=1.0\linewidth]{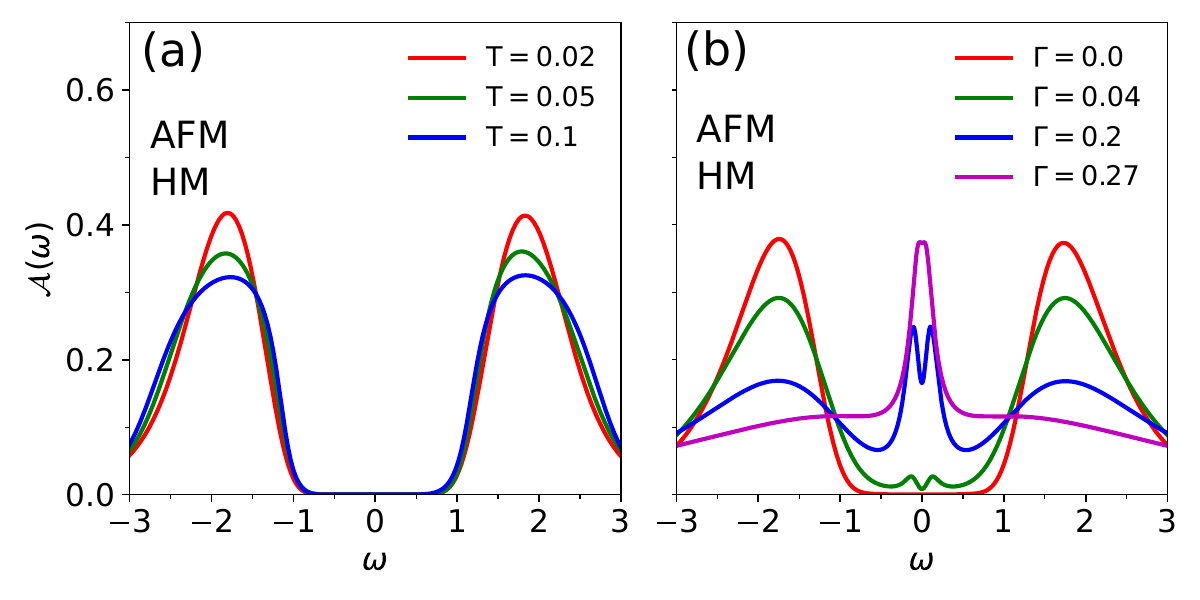}
\end{center}
\caption{Comparison of the evolution of the total (i.e. spin averaged) local DOS 
as a function of $T$ at fixed $\Gamma=0.0$ (a)
and of $\Gamma$ at fixed $T=0.01$ (b), at interaction $U$=4.0, for the AFM-HM across the AFM to PM. 
The line colors correspond to the spots along the $m_s$ curve of Fig.\ref{MS_T_Gamma}.}\label{DOS_AFM_U4.0}
\end{figure}

In contrast, at large $U=4.0$, the evolution of the local DOS with increasing $T$ and $\Gamma$ 
differs qualitatively from the low $U$ case. 
The DOS shown in the panel (a) of Fig.~\ref{DOS_AFM_U4.0} hardly changes as $T$ is increased,
even though the system undergoes an AFM to PM transition \cite{Rozenberg_PRL_optic,Camjayi_PRBRC}. In this case, therefore, instead of an 
insulator-to-metal transition we have a insulator-to-insulator one. 
At this large $U$, the charge degrees of freedom are frozen, which is signaled by the large Mott gap 
that can be observed. Hence, only the spin degrees of freedom respond to increasing $T$, which 
leads to the suppression of the (antiferro-)magnetic polarization, but does not affect the 
charge gap in the DOS, as typical in the Heisenberg regime. 

In contrast, the evolution as a function of $\Gamma$ for low $T$=0.01 is qualitatively different. 
We observe that in this case, the local DOS results do reveal an insulator-metal transition. 
Nevertheless, this transition is also qualitatively different from that of the lower $U$ case.
The main feature that emerges in the DOS upon the increase of $\Gamma$ is the emergence of 
a two in-gap peak structure around zero frequency, within the Mott-Hubbard band gap.
We can understand these two features in terms of the associated impurity problem. Since
the state is AF, the spin symmetry is broken into a N\'eel type of order into two sublattices. 
Thus, at a given
lattice site, the one type of spin has a larger occupation than the opposite, and vice-versa
on the neighboring site. Thus, one sublattice will have a, say, positive magnetic moment with
a higher occupation of the spin-up electrons, while the other sublattice will have the opposite.
Thus, each associated impurities, one corresponding at either lattice, will realize a different
state. Of course, there is up-down or particle-hole symmetry relating those two states, due
to the N\'eel order.

The crucial point is, nevertheless, that both impurities are in an environment that is not
fully gapped as before, but has low energy states due to the finite $\Gamma$. This can be
directly seen from the hybridization term associated to $\Gamma$ in Eq.\ref{self-cons2}. 
The consequence for this is that there will be a small Kondo peak emerging near
the Fermi energy in the solution of the associated impurity. Since we have two
inequivalent impurity sites, one for each sublattice, they lead to the two small
quasi-particle peaks that appear in the local DOS in Fig.\ref{DOS_AFM_U4.0}.
As we discussed before, the contribution to the effective bath associated to 
$\Gamma$ does not renormalize and remains the same for both spin projections and
for both sublattices. As $\Gamma$ increases we see that the strength of the Kondo
peaks increases. Eventually, the screening becomes so strong that the magnetic order
collapses as $m_s \to 0$ at $\Gamma_N$. At that point, there is a sole quasiparticle peak
in the local DOS since the spin symmetry is restored, and the system has gone through an 
insulator to strongly correlate metal transition.

\begin{figure}[h!]
\begin{center}
\includegraphics[width=1.0\linewidth]{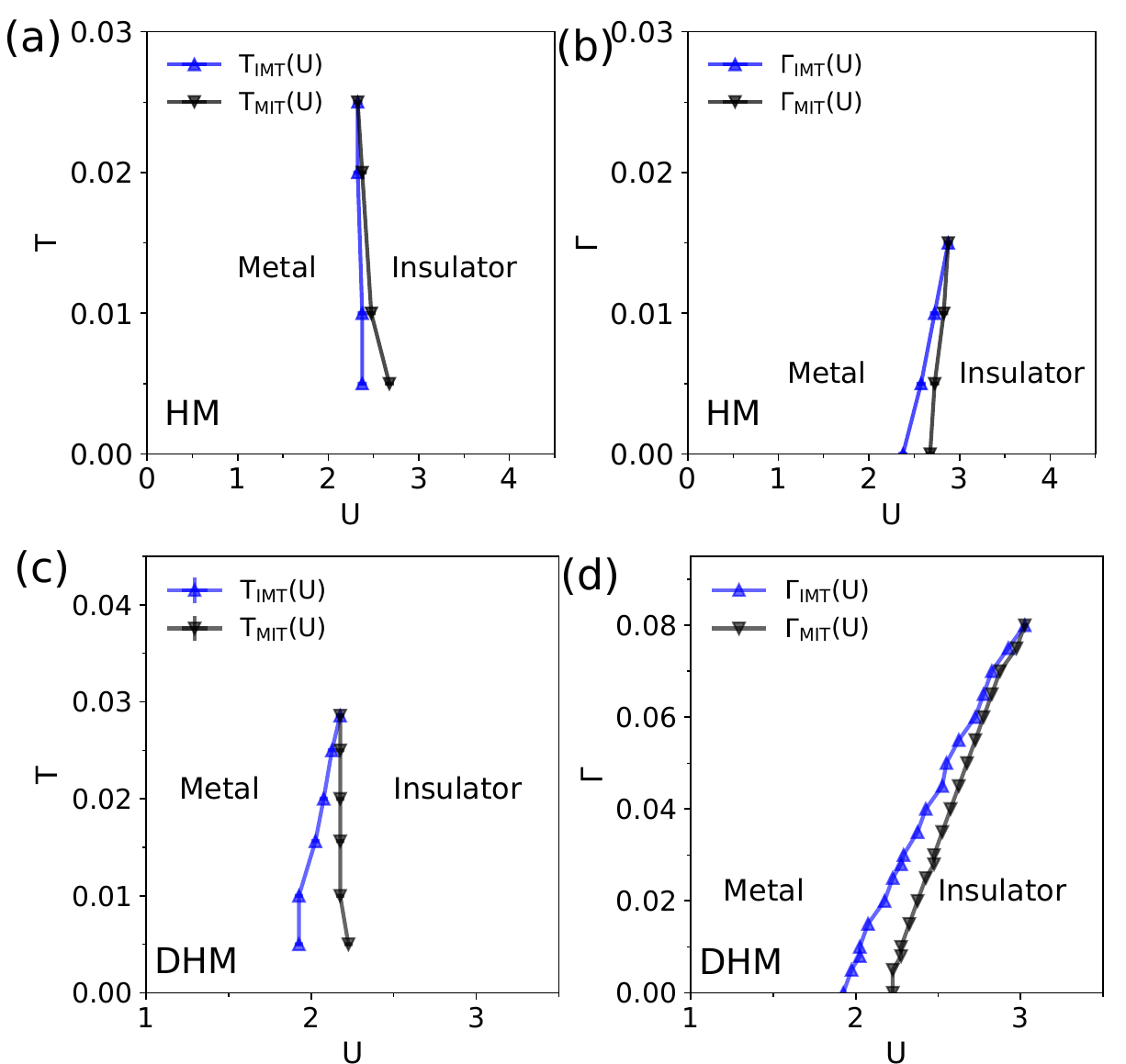}
\end{center}
\caption{(a) $T$-$U$ phase diagram at $\Gamma=0.0$ (b) $\Gamma$-$U$ phase 
the diagram at $T=0.01$ of the half-filled HM with PM symmetry. (c) $T$-$U$ phase diagram 
at $\Gamma=0.0$ (d) $\Gamma$-$U$ Phase diagram at $T=0.005$ of the DHM. 
In the $T$-$U$ ($\Gamma$-$U$) phase diagram, while increasing $U$ at a given $T$ ($\Gamma$), 
the system goes from metal to insulator at $T_{MIT}$(U) ( $\Gamma_{MIT}$(U)) 
and while decreasing $U$ system goes from insulator to metal at $T_{IMT}$(U) 
($\Gamma_{IMT}$(U)). The triangular regions denote the coexistence of solutions
which is consistent with the first order nature of the experimentally observed
transitions.{}}
\label{phase_diagram_HM_DHM}
\end{figure}

\subsection{Hubbard and Dimer Hubbard models in the paramagnetic state}

In previous sections, we have compared the effects of $T$ and $\Gamma$ on the AFM-PM transition of the HM, which are insulator-to-metal
at low $U$ and insulator-to-insulator at large $U$. We observed that the
effects of bath coupling can be considered qualitatively similar at low interaction $U$, but become substantially different 
at large $U$. 
We point out that both, the $T-$ and $\Gamma-$driven transitions, start out from an insulator state at all $U$, which is
gapped. While this first study was motivated by the experimental results on the transition in the V$_2$O$_3$ heterostructure,
it may be interesting to complete the picture to consider the case of the effect of $\Gamma$ in the well studied
paramagnetic MIT in the Hubbard model as well \cite{DMFTRMP}. 
In fact, as in this transition the AF order is absent, the system starts in a correlated metal without a gap.
So in this section we shall consider the interesting question of whether $\Gamma$ may also play an analogous role as $T$ does,
but this time within paramagnetic states. We shall consider two paradigmatic strongly correlated MIT, that of the HM and
the DHM one.

\subsubsection{Phase Diagrams}
The MIT in the Hubbard model within the paramagnetic phase is by now a well known a paradigm of a strongly correlated phenomena \cite{DMFTRMP}.
The phase diagram in the $T-U$ plane shows a metallic region at low $U$ and low $T$, a Mott insulator at high $U$ and low $T$, and a bad metal
at higher $T$ in the intermediate $U$ regime. One of the most interesting features is the existence of a coexistence region
in between the correlated metal and the Mott insulator, which gives a first order character to the transition. These well known results
are shown in the phase diagram at the panel (a) of Fig.~\ref{phase_diagram_HM_DHM}.
Following a similar strategy as before, we compared the $T-$driven behavior with that of $\Gamma$. The results are shown in the panel (b)
of the same figure. We strikingly observe that the behavior of the two parameters is qualitatively similar and the main feature of 
a triangular coexistence region is preserved. However, there is a qualitative difference, as it can be observed that the tilting of the
triangular region has changed.

To gain further insight we turned to the other basic model that we are considering in the present work, the DHM, which may be relevant,
as a first approximation, to the experiments in VO$_2$ hetero-structures. Indeed, the DHM also presents a coexistence region in the
$T-U$ phase diagram within the paramagnetic phase \cite{Oscar2017,Oscar2018,stinson2018imaging}. The phase diagram obtained with CT-QMC
simulations is shown in the panel(c) of Fig.~\ref{phase_diagram_HM_DHM}. We observe in this case that the triangular region
is tilted to the right, which is a common feature of cluster DMFT models, and the DHM can be considered as the simplest of those 
ones \cite{Oscar2017}. We explored by numerical simulations the fate of the triangular region at the lowest $T$ and increasing $\Gamma$.
The obtained phase diagram is shown in the panel (d) of the figure. Once again, we observe the feature that the
coupling to the electronic reservoir seems to play a similar role as the $T$, as now even the triangular coexistence region shows
a similar tilt.

\begin{figure}[t!]
\begin{center}
\includegraphics[width=1.0\linewidth]{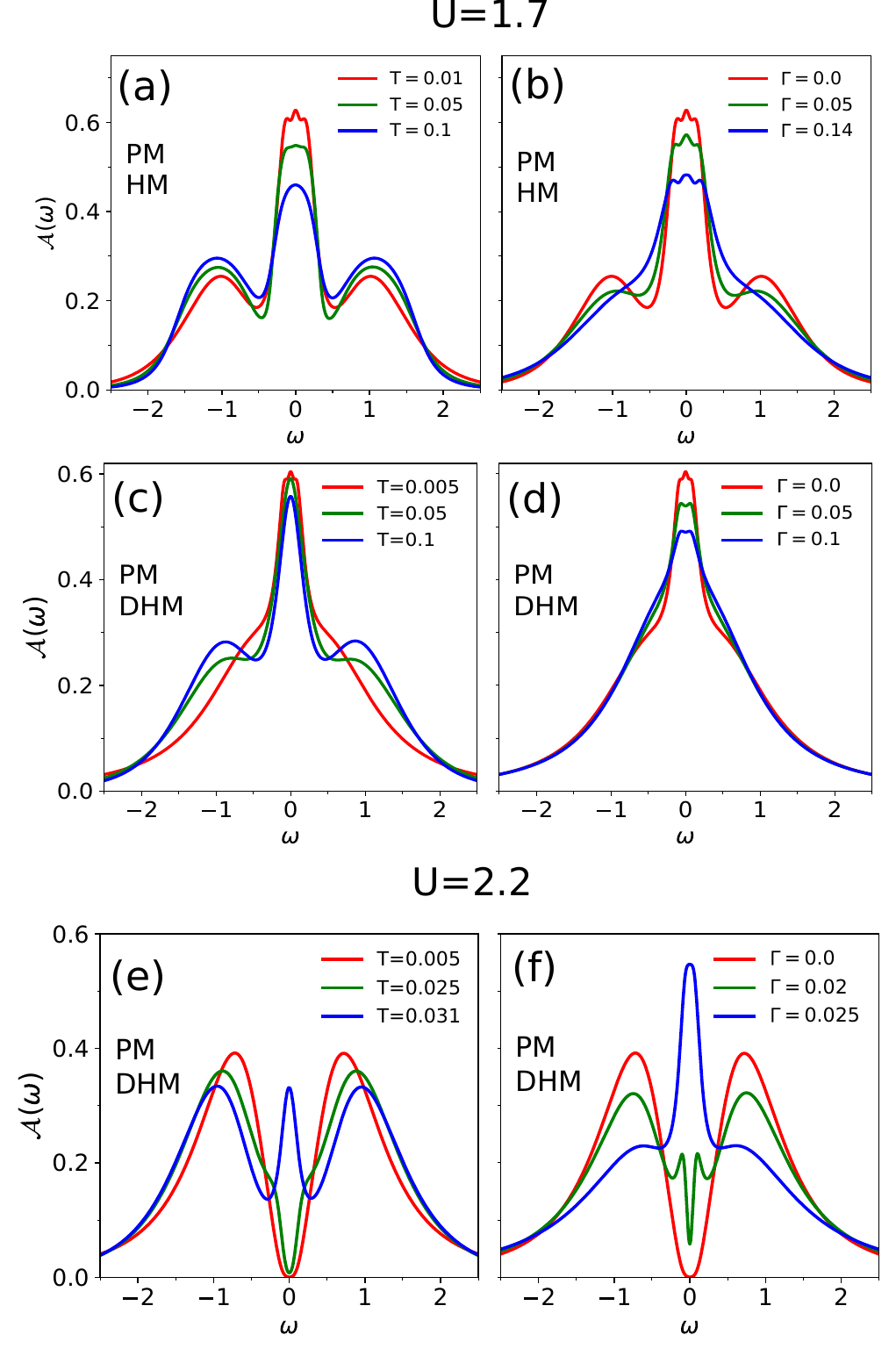}
\end{center}
\caption{Side-by-side comparison of the evolution of DOS as a function of $T$ with fixed $\Gamma=0.0$(a, c and e) and $\Gamma$ with fixed $T$(b, d and f) in weak interaction region for PM-HM and PM-DHM. (a and b) and (c and d) are for PM-HM and PM-DHM at $U$=1.7, respectively. (e) and (f) is for  PM-DHM at $U$ = 2.2. \cite{footnote} }\label{DOS_AtSmallU}
\end{figure}

\subsubsection{Density of States}
To understand further the effect of $T$ and $\Gamma$ at small $U$ and at large $U$, we study the systematic changes of DOS for 
the two models at small and large $U$. We begin with the results at low $U$ which are obtained by CT-QMC plus maximal entropy analytic
continuation. In Fig.\ref{DOS_AtSmallU} we show the DOS at the small $U$=1.7 (i.e metallic side of the transition) for the
HM and DHM.  
Interestingly, we observe that increasing both, $T$ and $\Gamma$ have a qualitatively similar effect in the DOS. They can both control the intensity of the quasiparticle peak, however, increasing T provokes a more significant transfer of weight toward the Hubbard bands with respect to $\Gamma$.

\begin{figure}[t!]
\begin{center}
\includegraphics[width=1.0\linewidth]{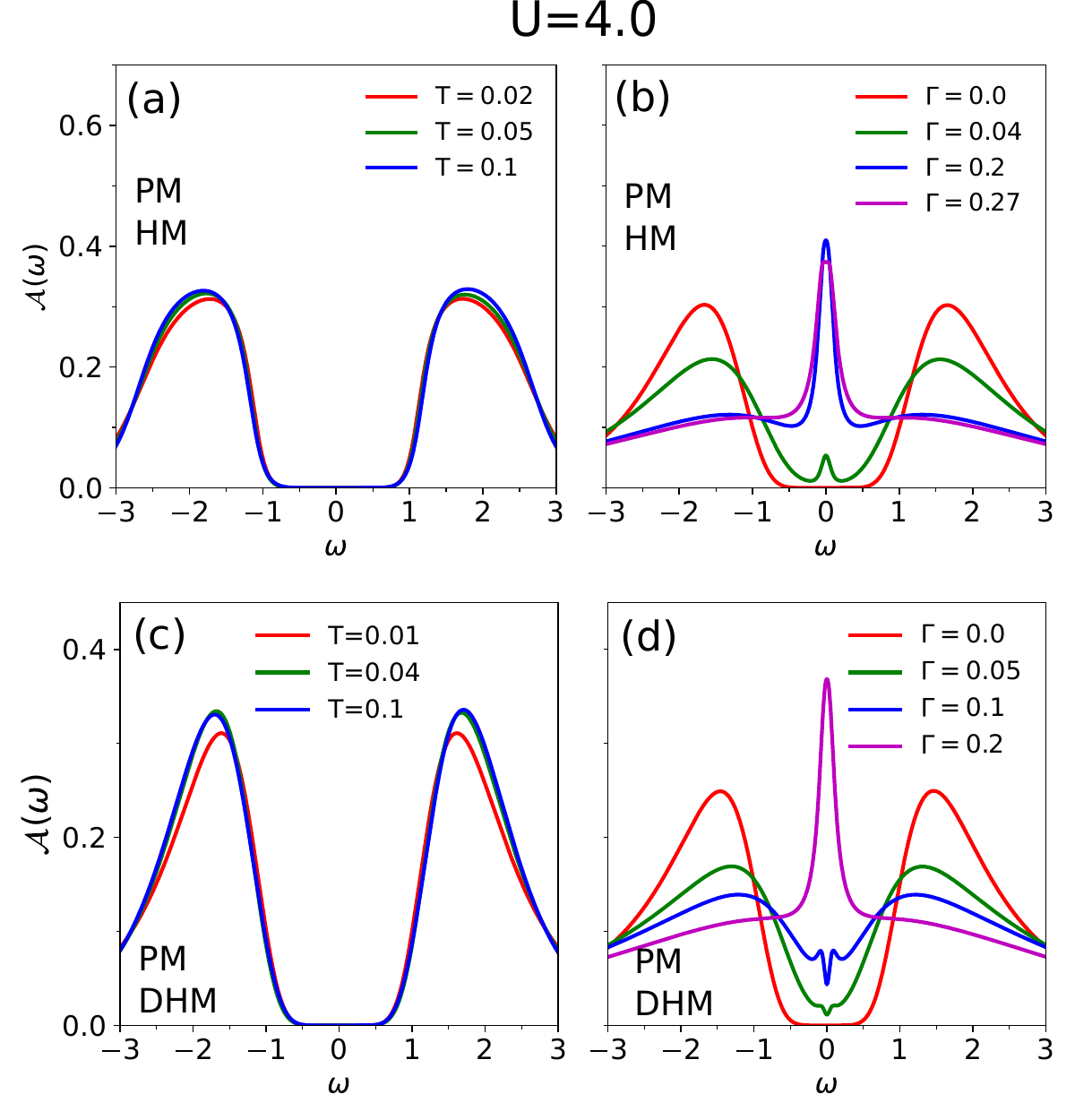}
\end{center}
\caption{(Side-by-side comparison of the evolution of DOS as a function of $T$ (a and c) and $\Gamma$ (b and d) at strong interaction $U=4.0$ for HM and DHM. (a) and (b) DOS at various $T$ for fixed $\Gamma=0.0$ and at various $\Gamma$ for fixed $T$, respectively, of HM without AFM symmetry. (c) and (d) shows DOS of DHM at $U$ = 4.0 at various $T$ with fixed $\Gamma=0.0$ and at various $\Gamma$ for fixed $T$=0.05 respectively.}\label{DOS_HM_DHM_U4}
\end{figure}

We also explore a stronger correlated state for the DHM, namely at $U$=2.2 within the coexistence region, which is
considered relevant for the VO$_2$ compound \cite{Oscar2017}. In this case, we start from the insulator state
within the coexistence, which as a function of $T$ exhibits a first-order insulator-to-metal transition in qualitative 
agreement with the mentioned compound.
The results for the evolution of the DOS are shown in the panel (e) of Fig.~\ref{DOS_AtSmallU}, where
we observe the metallization of the correlated insulator.
In the panel (f), we report the behavior as a function of $\Gamma$ for the same starting state.
Quite strikingly, we observe a similar behavior. The gap is filled and a prominent quasi-particle peak emerges
at the Fermi energy, reminiscent of Kondo physics.
Nevertheless, we should also mention that a relatively small difference seems to be present in the evolution of
the respective spectra, since at intermediate values of $\Gamma$ we observe two small quasi-particle peaks at the
inner edges of the Hubbard bands, which are not present (or very subtle) in the $T$-driven case.
Another difference is that in the correlated metal state driven by $\Gamma$ the Hubbard bands have relative
smaller spectral intensity as compared to the $T$-driven transition, which is similar to the effect noted at smaller $U$.

For completeness, we also considered the large $U$ region, where both systems are deep in the Mott insulator phase.
We adopt the relatively large value $U$=4.0. The results are shown in Fig.\ref{DOS_HM_DHM_U4},
where we track changes in the DOS for the two models (HM and DHM) as a function of $T$ and $\Gamma$.  
With increasing temperature at $\Gamma$=0 on the left hand side we observe that the DOS hardly changes for both, HM and DHM. 
Which is natural since the excitation gap at $U$=4 is much larger than $T$. However, in striking contrast, the respective DOSs 
at low $T$ and increasing $\Gamma$, shown on the right hand of the figure, develop a prominent quasi-particle peak at the Fermi energy. 

The interpretation of these behavior also is rooted in the Kondo effect. Indeed, the large value of $U$ create a strong local magnetic
model, which at $\Gamma$=0 is essentially an uncoupled spin, which is a peculiarity of the Mott paramagnet \cite{DMFTRMP}. 
When the coupling with the electronic bath is switched on, the bath electrons immediately screen the local spins forming Kondo
resonances. We may note that a emergence of a Kondo peak in a large Mott gap was reported in systems with a narrow correlated band 
coupled with a wide conduction band \cite{Medici} and in the hetero-structure of a metal coupled to a Mott insulator\cite{HM-M1,HM-M2}. 

Comparing Figs.~\ref{DOS_AtSmallU} and \ref{DOS_HM_DHM_U4}, and consistent with the Kondo origin of the phenomenon, 
we observe that the quasi-particle-peak weight increases with $\Gamma$ at constant $U$,
but decreases with increasing $U$ at constant $\Gamma$. 
At large $U$, both the HM and the DHM can be driven across an MIT by the intensity of the coupling to the reservoir, $\Gamma$,
but not by $T$. We may also point out that at large $U$ we also observe in the DOS of the DHM the two small edge quasiparticle peaks
that develop as the gap closes.
\\



\begin{figure}[h!]
\begin{center}
\includegraphics[width=1.0\linewidth]{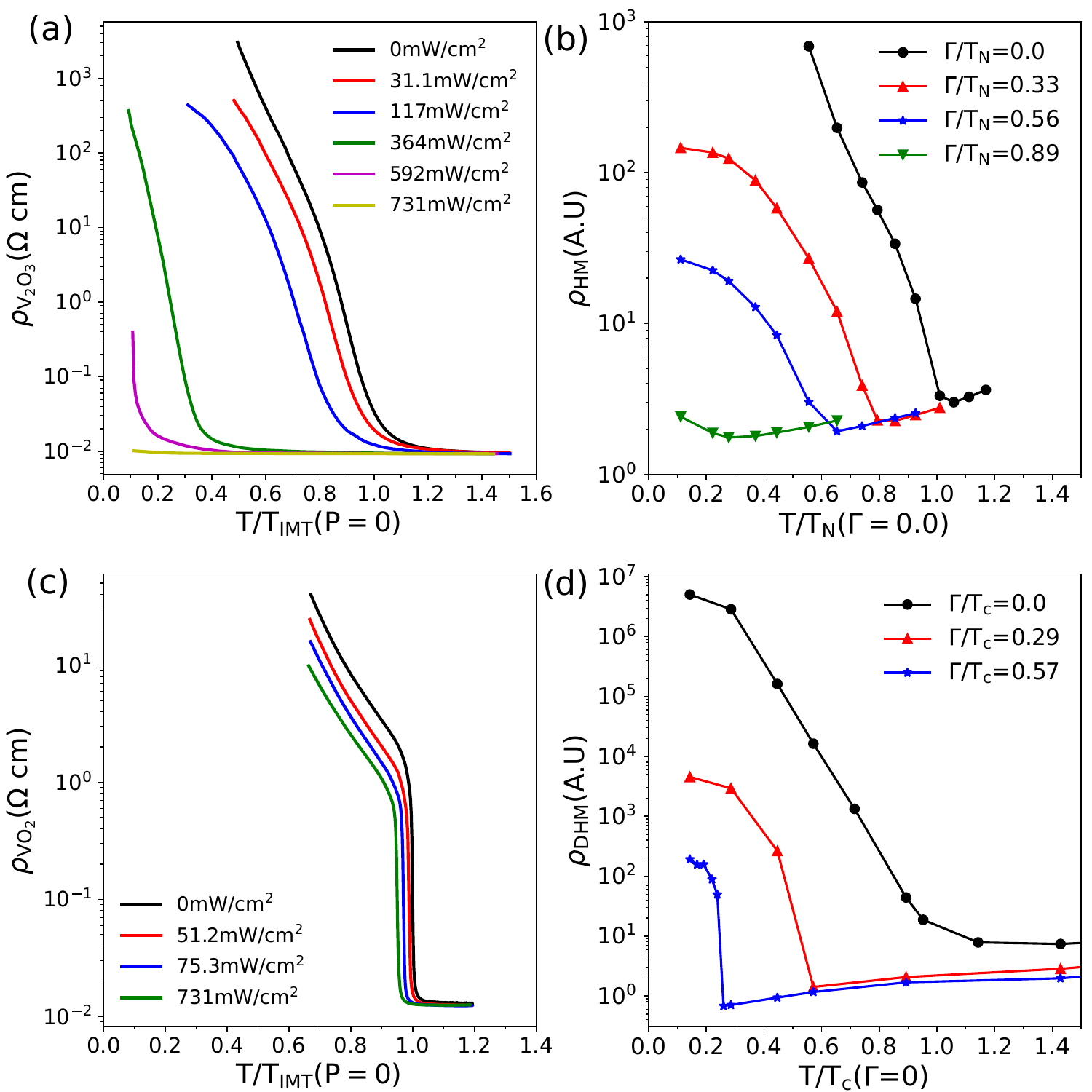}
\end{center}
\caption{ Comparing temperature-dependent experimental resistivity with theoretically calculated resistivity for various light intensities (P) on CdS/V$_2$O$_3$ and CdS/VO$_2$. (top-left) and (bottom-left) experimental resistivity ($\rho$) for CdS/V$_2$O$_3$ and CdS/VO$_2$, respectively\cite{navarro2021hybrid}. (top-right) and (bottom-right) variation of $\rho$ of HM-AFM at $U$ = 1.7  and  DHM at $U$=2.2, respectively, as a function of $T$ for various values of $\Gamma$. We scale the $T$ with the $T_{MIT}$ of CdS/V$_2$O$_3$ with no light on CdS, which is 148 K.}\label{experiment}
\end{figure}

From the systematic study of the model done so far, with and without magnetic order and at low and large values of $U$, 
we observe that there is an emergent pattern. Namely, we can condense our results noting that to zero order, the effect of $T$ and
$\Gamma$ are qualitatively similar when the interaction $U$ is relatively low. One may be tempted to say that this is in the metallic state
and the reason is that the local coupling to the incoherent electron bath plays an analogous role as thermal scattering. This is
qualitatively true, however, the argument has even more general validity, since in the AFM {\it insulator} at low $U$, both $T$ and $\Gamma$
also have a qualitatively similar effect. Interestingly, in a recent paper \cite{BROUET}
the qualitative difference of the behavior of the AFM Hubbard model at low and large $U$ was investigated. The results showed that
in the former case the system is described by a Slater AFM insulator where the bands are splitted by a gap but retain their coherent (i.e.
quasiparticle) character. In contrast, at large $U$ the system is a Heisenberg AFM insulator where the (Hubbard) bands are fully incoherent.
Hence, in the low $U$ AFM the coupling with the local electron reservoirs disturb the coherent character of the quasi-particle propagation,
pretty much as $T$ does.

For completeness, we have constructed a full 3D $T-U-\Gamma$ phase diagram of the HM and DHM where the global behavior
in parameter space may be appreciated and is presented in the Supplementary Material.  
\\

\section{Comparison with experiments}

We now compare our results with the experimentally observed effect of light illumination on the MIT of vanadate
thin films with a deposited layer of photoconductor CDS. 
The panel (a) of Fig.~\ref{experiment} shows the variation of resistivity ($\rho$) of CdS/V$_2$O$_3$ as a function of $T$ 
for various values of illumination intensity $P$ (that we qualitatively identify with $\Gamma$ see Section II). 
The system shows a sharp resistance change across the insulating to metal transition.  
Significantly, we also note that the transition temperature decreases with increasing illumination intensity. 
The system's $T$$_{MIT}$ is driven down to zero temperature for high enough light intensity. 
We may favourably compare those experimental results with our calculations of the variation of resistivity
$\rho(T)$ in the AFM-HM coupled with the electronic reservoir at $U$=1.7 for various values of $\Gamma$, that we show 
in the panel (b) of Fig.~\ref{experiment}. For better comparison, we scaled the $T$ with N\'eel 
temperature ($T_N$) at $\Gamma$=0, which is  0.09. 
The resistivity of our model calculation also shows a sharp decrease of resistivity across MIT, 
which is associated with the AFM to PM transition. 
Importantly, one can also note that transition temperature decreases with increasing $\Gamma$. 
Thus, AFM Hubbard model coupled with an electronic reservoir can qualitatively capture the 
suppression of $T_{MIT}$ observed in CdS/V$_2$O$_3$. While the simple one band Hubbard model may seem
a simplistic model of V$_2$O$_3$, and in many aspects it is, it is worthy to note that the same model
for the same value of the parameter $U$ was adopted to successfully account for non-trivial
large negative magneto-resistance effects observed in thin films of the same compound \cite{V2O3_HM}

We now turn to the experiments done on VO$_2$.
The panel (c) of Fig.\ref{experiment} shows the experimental resistivity of CdS/VO$_2$ for 
various values of illumination intensity. The system shows a sharp insulating to metal transition as a function 
of temperature with sharp resistance change at $T_{MIT}$. Similarly as before, we scale the $T$ with the $T_{MIT}$=308K 
of CdS/VO$_2$ with no light on CdS. Rather surprisingly, there is also a reduction in $T_{MIT}$ upon illumination
as seen in CdS/V$_2$O$_3$, but the magnitude of the effect for the same light intensity is much less sensitive.
We turn to the DHM and set the parameter as were adopted in Ref. \cite{Oscar2017,stinson2018imaging}, namely 
$t_\perp$= 0.3 and $U$=2.2 (near the coexistence region) at half-filling ($\mu$ = $U$/2) where that model successfully accounted
for some experimentally observed features in VO$_2$. 
The panel (d) of Fig. \ref{experiment} shows the calculated resistivity $\rho(T)$ of the DHM 
coupled with the electronic reservoir for various values of $\Gamma$. 
For better comparison, we scaled $T$ with the critical temperature $T_c$=0.03 at $\Gamma=0.0$. 
One can note that transition temperature decreases with increasing $\Gamma$ as observed for CdS/VO$_2$ and in the previous
case. However, the effect seems too strong as compared to the experimental findings.
We have tried varying different parameters in different regions of the phase diagram, but did not find any better agreement.

Since the comparably simple Hubbard model has accounted for V$_2$O$_3$ and in view of past success of the DHM model to account
for aspects of VO$_2$, we should try to speculate on the origin of the discrepancy. One natural missing ingredient in the model
may be that the structural transition may play a significant role. VO$_2$ has a larger structural transition that is
concurrent with the MIT than V$_2$O$_3$, so its relative effect may be more significant \cite{entropy}. 
In fact, one may argue that since a structural transition can be understood as due to the 
condensation of phonon fluctuations, therefore a purely electronic reservoir may not be effective in driving the system sufficiently.
Therefore the effect is still present but in a much smaller magnitude as the MIT is always arrested by the inability
of the CdS to promote phononic excitations.


\section{Conclusion}\label{ConclusionAndDiscussion}
 
 We study the effect of the
 photoconductor CdS on two strongly correlated vanadate thin films. 
 The theoretical modeling that we introduced consists of a collection of incoherent electronic reservoirs 
 coupled at every lattice site of a Hubbard and a Dimer Hubbard model. 
 We note that besides the motivation to account for the experimental observations, these models are also interesting from a
 fundamental physics point. Indeed, the coupling with electronic reservoirs is a required feature in order to describe electrons on lattices, driven out-of-equilibrium by external electric fields. The currents that develop produce heat that
 need to be evacuated via reservoirs to allow the system to thermally equilibrate \cite{diaz2023electrically}.
 
The coupling strength of those models with the electronic reservoirs is characterized by a parameter $\Gamma$, 
which we argued may play the similar role as the illumination power. 
These models are solved using DMFT with HYB-CTQMC as an impurity solvers. 

From the systematic study we observed that $\Gamma$ and $T$ played a qualitatively similar role, so long the
underlying electronic structure could be described with quasi-particles, which is at low to moderate $U$, regardless
whether the system is a metal or an insulator. We trace that to the fact that the local coupling with the reservoir
produces a source of electronic scattering, similarly as the $T$ does. 

This observation provided a natural explanation to the dramatic suppression of the transition temperature in V$_2$O$_3$,
where the insulator state is due to the opening of an antiferromagnetic gap, i.e. of electronic origin. Hence the
electronic reservoir was efficient to suppress it.
In contrast, the agreement was not so successful for the VO$_2$ experiments, which show a significant smaller effect
of reduction in the transition temperature. We argue that this points to a prominent role of the structural transition,
which is couple to lattice vibration degrees of freedom \cite{MFAA}, hence the electronic reservoir is less efficient.

It will be interesting in future work to consider the extension of our hybrid model to other strongly correlated models,
such as the double exchange model and the Periodic Anderson model, which are relevant for other material compounds and
may open the way to new forms of control of strongly correlated phenomena.

\section{Acknowledgements}
We acknowledge support from the French ANR
“MoMA” project ANR-19-CE30-0020. LF is supported by 
the Quantum Materials for Energy Efficient Neuromorphic 
Computing, an Energy Frontier Research Center funded 
by the US Department of Energy, Office of Science, Basic 
Energy Sciences under Award DE-SC0019273. MC is supported 
by French ANR ``Neptun" ANR-19-CE30-0019-01. We thank H. Navarro of UCSD for providing the 
figure with experimental data and for useful discussions.
 
\appendix
\section{DC conductivity} \label{conductivity_equation}
  The Drude conductivity of HM on Bethe lattice was calculated using Eqn.
\begin{align}
\Re\sigma_{HM}&(\omega=0)
= 4\int d\epsilon \rho(\epsilon) \int d\omega 
\left(- \frac{\partial f(\omega)}{\partial \omega}\right)_{\omega=0} \times \\ \nonumber
&\left(\frac{D^2-\epsilon^2}{3}\right)\left(\mathcal{A}_{\uparrow}(\epsilon,\omega)\mathcal{A}_{\downarrow}(\epsilon,\omega)+ \mathcal{A}^2_{off}(\epsilon,\omega) \right)
 \end{align}
 
 where $\mathcal{A}_{\sigma}(\epsilon,\omega)=\frac{-1}{\pi} Im \frac{z_{\Bar{\sigma}}(\omega)}{z_{\Bar{\sigma}}(\omega)z_\sigma(\omega)-\epsilon^2}$ and $\mathcal{A}_{off}(\epsilon,\omega)=\frac{-1}{\pi} Im \frac{\epsilon}{z_{\Bar{\sigma}}(\omega)z_\sigma(\omega)-\epsilon^2}$ with $z_\sigma(\omega) =\omega+\mu -\Sigma_{\sigma}(\omega) - \Delta_\Gamma(\omega)$. $\frac{D^2-\epsilon^2}{3}$ is the velocity of electron of energy $\epsilon$ in Bethe lattice \cite{Chung}.

 The DC conductivity of DHM on Bethe lattice was calculated using Eqn.\cite{ocampo2017study}
\begin{align}
\Re\sigma_{DHM}&(\omega=0)= 2\int d\epsilon \rho(\epsilon) \int d\omega 
\left(- \frac{\partial f(\omega)}{\partial \omega}\right)_{\omega=0} \times\\ \nonumber
&\left(\frac{D^2-\epsilon^2}{3}\right)\left(\mathcal{A}^2_{AB}(\epsilon,\omega)+\mathcal{A}^2_{B}(\epsilon,\omega)\right)
 \end{align}\label{conductivity_DHM}
 
 Where $\mathcal{A}^2_{AB}(\epsilon,\omega)$ and $\mathcal{A}^2_{B}(\epsilon,\omega)$ are DOS of Anti-bond and Bond respectively.

\section{Metal-Insulator Coexistence regions}

We display here the metal-insulator coexistence region in the full $T-U-\Gamma$
phase diagram. Notice the difference in the tilting between the Hubbard (HM) and Dimer Hubbard (DHM) models by cutting $T-U$ planes.
\begin{figure}[htbp!]
\begin{center}
\includegraphics[width=0.6\linewidth]{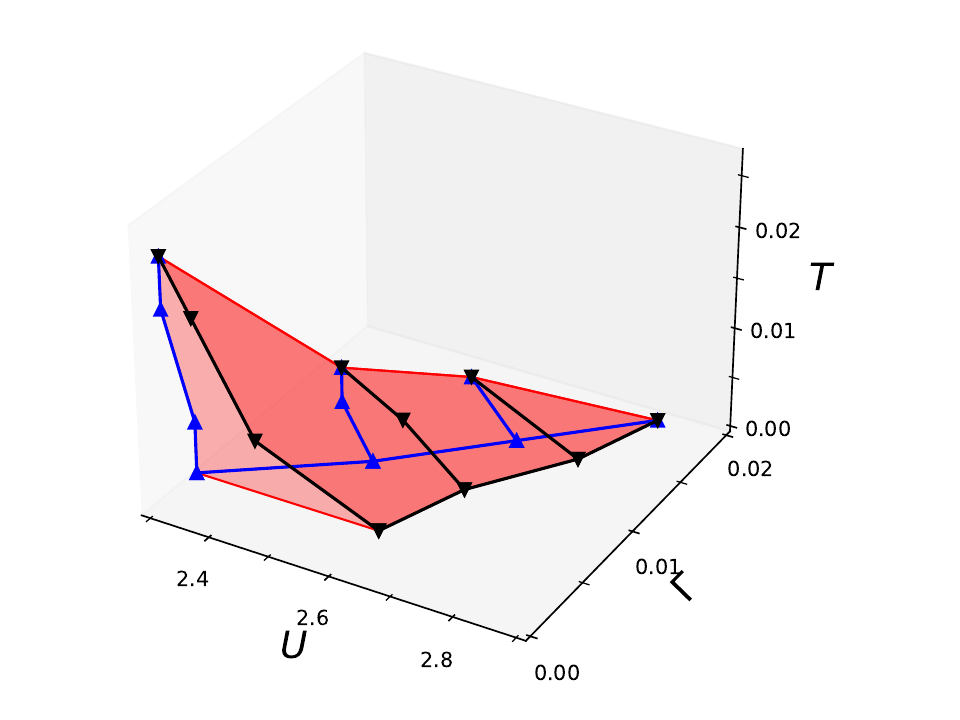}
\end{center}
\label{HM11}
\caption{$T$-$\Gamma$-$U$ phase diagram of HM at half filling. The coexistence region (colored region) in the $T-U$ plane of the phase diagram shifted toward the right as one increases $\Gamma$ and T$_c$ decreases with increasing $\Gamma$}
\end{figure}

\begin{figure}[htbp!]
\begin{center}
\includegraphics[width=0.6\linewidth]{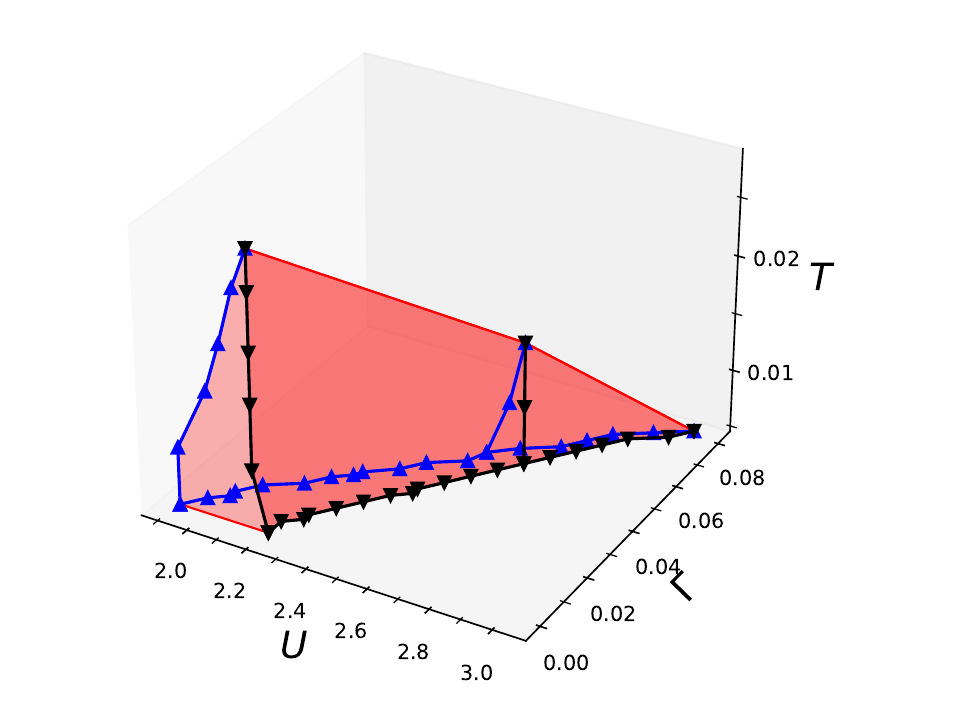}
\end{center}
\label{DHM11}
\caption{ $T$-$\Gamma$-$U$ phase diagram of DHM at half filling. The coexistence region (colored region) in the $T-U$ plane of the phase diagram shifted toward the right as one increases $\Gamma$ and T$_c$ decreases with increasing $\Gamma$}
\end{figure}

\pagebreak
\newpage
,\clearpage
\bibliographystyle{unsrtnat}
\bibliography{DHM_photo_doped2}{}
\end{document}